\documentclass[10pt]{article}
\usepackage[sort&compress]{natbib}
\usepackage{graphicx}
\usepackage[utf8]{inputenc}

\usepackage{url}
\usepackage{times}
\usepackage{placeins} 
\usepackage{multicol}
\usepackage[footnotesize,labelfont=bf,labelsep=period,justification=raggedright]{caption}

\newcommand{\nspace}[1]{\setlength{\baselineskip}{#1\spacing}}

\usepackage[small,compact]{titlesec}

\makeatletter

\newenvironment{figurehere}
{\def\@captype{figure}}
{}
\makeatother

\title{Phenotypic robustness can increase phenotypic variability after non-genetic perturbations in gene regulatory circuits}
\author{Carlos Espinosa-Soto$^{* 1,2}$
  \and
   Olivier C. Martin$^3$
    \and 
     Andreas Wagner$^{1,2,4}$
         }

\date{}           

\begin{document}
\maketitle

{\vspace{-2mm}
{\small
 $^1$ University of Zurich, Dept. of Biochemistry, Bldg. Y27 Winterthurerstrasse 190 CH-8057 Zurich, Switzerland
 
 $^2$ The Swiss Institute of Bioinformatics. Quartier Sorge, Batiment Genopode, 1015 Lausanne, Switzerland
 
 $^3$ INRA, UMR 0320 / UMR 8120 G\'en\'etique V\'eg\'etale, F-91190 Gif-sur- Yvette, France
 
 $^4$ The Santa Fe Institute, 1399 Hyde Park Road, Santa Fe, NM 87501, USA
 
 \vspace{1mm}
 
\parindent 0em
 Email: C. Espinosa-Soto$^*$ -- c.espinosas@gmail.com; O.C. Martin -- olivier.martin@u-psud.fr ; A. Wagner -- aw@bioc.uzh.ch

 $^*$ Corresponding author
 }
 \parindent 2em

\begin{abstract}
Non-genetic perturbations, such as environmental change or developmental noise, can induce novel phenotypes. If an induced phenotype confers a fitness advantage, selection may promote its genetic stabilization. Non-genetic perturbations can thus initiate evolutionary innovation. Genetic variation that is not usually phenotypically visible may play an important role in this process. Populations under stabilizing selection on a phenotype that is robust to mutations can accumulate such variation. After non-genetic perturbations, this variation can become a source of new phenotypes. We here study the relationship between a phenotype's robustness to mutations and a population's potential to generate novel phenotypic variation. To this end, we use a well-studied model of transcriptional regulation circuits. Such circuits are important in many evolutionary innovations. We find that phenotypic robustness promotes phenotypic variability in response to non-genetic perturbations, but not in response to mutation. Our work suggests that non-genetic perturbations may initiate innovation more frequently in mutationally robust gene expression traits. 
\end{abstract}

\begin{multicols}{2}
\section*{Introduction}
Two main perspectives exist about the origin of evolutionary innovations. The orthodox ``genotype-first'' perspective emphasizes the role of mutations in the production of new phenotypes. In this perspective, mutations produce individuals with novel phenotypes whose frequency in a population may increase through natural selection. The heterodox ``phenotype-first'' perspective \citep{West-Eberhard1989,Pigliuccietal2006,Hall2001,Newmanetal2006,Moczek2007,West-Eberhard2003,Palmer2004,Priceetal2003} emphasizes the role of non-genetic perturbations, such as exposure to different temperatures, diets, or biotic interactions. Non-genetic perturbations also comprise fluctuations in an organism's internal ``microenvironment'', such as gene activity changes caused by noisy gene expression \citep{McAdamsyArkin1997,Elowitzetal2002,Rajetal2010}.

The phenotype-first perspective is based on the observation that organisms often have highly plastic phenotypes. That is, the same genotype has the potential to produce different phenotypes depending on non-genetic influences. Thus, a perturbation can trigger a plastic phenotypic response in some individuals of a population. If the resulting novel phenotype provides a benefit to its carrier, it facilitates survival. Then, selection may increase the frequency of new or already existing genetic variants that exaggerate, refine or ``stabilize'' this phenotype by making it independent of non-genetic factors. Waddington coined the term genetic assimilation for this stabilization process \citep{Waddington1953}. 

Increasing amounts of evidence support the importance for innovation of traits induced by non-genetic factors \citep[for dissenting opinions, see][]{Orr1999,deJongyCrozier2003}. First, theoretical work shows that assimilation {\it can} occur under broad conditions \citep{Wagner1996,Cilibertietal2007,SiegalyBergman2002,Wagneretal1997,Rice1998,Lande2009,Masel2004,Espinosa-Sotoetal2010}.
Second, laboratory evolution experiments show that assimilation {\it does} occur \citep{Waddington1953,Waddington1956,Eldaretal2009,RutherfordyLindquist1998,SuzukiyNijhout2006}. Third, studies in natural populations suggest that genetic assimilation of traits induced by non-genetic factors is not rare \citep{West-Eberhard2003,PigliucciyMurren2003,Palmer2004}. For example, taxa with genetically determined dextral or sinistral morphologies are frequently derived from taxa in which the direction of the asymmetry is not genetically fixed, but where it is a plastic response \citep{Palmer1996pnas,Palmer2004}. This occurs for many traits, such as the side on which the eye occurs in flat fishes (Pleuronectiformes), and the side of the larger first claw in decapods (Thalassinidea) \citep{Palmer1996pnas}. Transitions like these indicate genetic assimilation of a direction of asymmetry originally induced by non-inheritable factors. More generally, traits where fixed differences among closely related species are mirrored by plastic variation within populations are good candidates for genetic assimilation. For example, amphibian traits, such as gut morphology \citep{Ledon-Rettigetal2008} and limb length and snout length \citep{GomezMestreyBuchholz2006}, follow this pattern. 

A system is robust to genetic or non-genetic perturbations if its phenotype does not change when perturbed. Mutational robustness and robustness to non-genetic perturbations are correlated with one another in many cases \citep{Cilibertietal2007,AncelyFontana2000,RutherfordyLindquist1998,deVisseretal2003,MeiklejohnyHartl2002,Lehner2010,Proulxetal2007}, although exceptions exist \citep{MaselySiegal2009,Cooperetal2006,FraserySchadt2010}. 

The ability to produce evolutionary innovation, is linked to the robustness of a biological system \citep{DraghiyWagner2009,Cilibertietal2007pnas,AncelyFontana2000,Wagner2005,Wagner2008nrg}. At first sight, robustness seems to hamper innovation. First, a mutationally robust system produces less phenotypic variation in response to mutations. It may thus not facilitate the genotype-first scenario \citep{DraghiyWagner2009,Cilibertietal2007pnas}. Second, a system robust to non-genetic factors shows little phenotypic plasticity. Thus, it may not support innovation under the phenotype-first scenario. However, the role of robustness in innovation is subtler than it may seem. This becomes evident when one considers how genotypes and their phenotypes are organized in a space of genotypes. 

Genotypes exist in a vast space of possible genotypes. Two genotypes are neighbors in this space if one can be transformed into the other by a single mutation. The distribution of phenotypes in genotype space shows some qualitative similarities for different kinds of systems, from RNA and protein molecules to metabolic networks and transcriptional regulation circuits. First, in these systems large sets of genotypes produce the same phenotype. Each of these sets can be traversed through single mutation steps that leave the phenotype unchanged. Such a set is also referred to as a neutral network or genotype network \citep{Schusteretal1994}. Second, different neighborhoods of the same genotype network contain genotypes with very different genotypes \citep{SchultesyBartel2000,Schusteretal1994,Wagner2008prsb,FerradayWagner2008,Cilibertietal2007pnas,RodriguesyWagner2009,LipmanyWilbur1991}. 

To understand how mutational robustness relates to a system's ability to produce evolutionary innovations, it is useful to distinguish between the mutational robustness of a genotype and that of a phenotype. A genotype $G_1$ is mutationally more robust than another genotype $G_2$, if $G_1$ is more likely to maintain the same phenotype than $G_2$ in response to mutation. By extension, a phenotype $P_1$ is mutationally more robust than $P_2$ if the genotypes that produce $P_1$ preserve $P_1$, on average, more often than the genotypes adopting $P_2$ preserve $P_2$ in response to mutations. Perhaps surprisingly, mutational phenotypic robustness can facilitate the production of novel phenotypes for RNA structure phenotypes \citep{Wagner2008prsb}. The reason is that genotypes with a more robust phenotype form larger genotype networks and have, on average, more neighbors with the same phenotype. A population of such genotypes encounters relatively few deleterious mutations that would slow its diversification and spreading through genotype space (while preserving its phenotype). It thus attains a higher genotypic diversity which translates into greater phenotypic variability in response to mutations, even though every single genotype may have access to fewer other phenotypes \citep{Wagner2008nrg}. 

This mechanism, although corroborated for RNA and protein structural phenotypes \citep{Wagner2008prsb,FerradayWagner2008} may not lead to increased phenotypic variability in all systems. The reason is that it depends on how many different and unique phenotypes the neighborhood of different genotypes contains, and on how rapidly populations can spread through a genotype network. In other words, it depends on the organization of genotype networks in genotype space, which may differ among different system classes. For example, a recent theoretical analysis suggests that the relationship between phenotypic mutational robustness and the potential to generate phenotypic variation through mutation need not even be monotonic \citep{Draghietal2010}. 

The above considerations pertain to phenotypic variability in response to mutations. One might think that phenotypic variability in response to non-genetic perturbations may behave similarly since robustness to mutations and to non-genetic factors are often positively correlated \citep{Cilibertietal2007,AncelyFontana2000,RutherfordyLindquist1998,deVisseretal2003,MeiklejohnyHartl2002,Lehner2010}. However, we show that this is not necessarily so for transcriptional regulation circuits, which exist on a higher level of organization than individual evolving molecules. Such circuits direct the production of specific gene activity patterns at particular times and places in the developing organism. Changes in the expression of their genes are involved in many evolutionary innovations \citep{Shubinetal2009,DavidsonyErwin2006}. We study a generic computational model of transcriptional regulation in which the genotypes correspond to the cis-regulatory interactions in a transcriptional circuit. The phenotypes correspond to the gene activity pattern a circuit produces. 

For this system, we show that high phenotypic robustness to mutations increases the number of novel expression phenotypes a circuit can produce in response to non-genetic perturbations. Thus, phenotypic robustness to mutation facilitates innovation under the phenotype-first scenario. 
It does so by allowing the accumulation of genetic variation that is not observed phenotypically under typical conditions, but that may be exposed after non-genetic perturbations \citep{deVisseretal2003,MaselySiegal2009,GibsonyDworkin2004,Masel2006}. 

\section*{Methods}
\subsection*{Model}

The model represents a regulatory circuit of $N$ genes, where each gene's activity is regulated by other genes in the circuit. The circuit's genotype is defined by a real-valued matrix ${\mathbf A} = (a_{ij})$, in which non-zero elements represent regulatory interactions between genes (Fig.~1a). An interaction ($a_{ij} \neq 0$) means that the activity of gene $j$ can either have a positive ($a_{ij} > 0$) or a negative ($a_{ij} < 0$) effect on the activity of gene $i$. We use $m$ to refer to the number of interactions in a given circuit, and $c$ to its interaction density, i.e. to the number of interactions $m$ divided by the maximum possible number of interactions $N^{2}$. A vector $s_t = (s_{t}^{(1)},...,s_{t}^{(N)})$ describes the activity state of the circuit at time $t$.

The activity of the genes in the circuit changes according to the difference equation 

\begin{equation}
 s_{t+\tau}^{(i)} = \sigma \left[\sum_{j=1}^{N} a_{ij}s_{t}^{(j)} \right]
\end{equation}
where $\sigma (x)$ equals -1 when $x < 0$, it equals 1 when $x > 0$, and it equals 0 when $x=0$.

Despite its level of abstraction, variants of this model have proven useful for studying the evolution of robustness in gene regulatory circuits \citep{Cilibertietal2007,Wagner1996,SiegalyBergman2002,MartinyWagner2008}, the effect of recombination on the production of negative epistasis \citep{Azevedoetal2006,MartinyWagner2009}, the evolution of modularity in gene circuits \citep{Espinosa-SotoyWagner2010} and the evolution of new gene activity patterns \citep{Cilibertietal2007pnas,DraghiyWagner2009}. Similar models have also been successfully used to predict the dynamics of developmental processes in plants and animals \citep{Mjolsnessetal1991,MendozayAlvarez-Buylla1998}.

We consider circuits that start their dynamics from a particular initial gene expression state $s_0$. One can view this initial state as being specified by factors external to the circuit, be they environmental factors, signals from adjacent cells, maternal regulators, or any genes ``upstream'' of the circuit. The phenotype is the stable gene activity pattern $s_\infty$ that a circuit attains when starting from $s_0$. We here focus on circuits producing such stable patterns and disregard circuits with more complex dynamics. We define a circuit's fitness by a function that increases steeply with the similarity of its phenotype $s_\infty$ to a reference activity state $s_\infty^{opt}$, that is considered the optimal phenotype in a given environment. We consider circuits that attain the same $s_\infty$ as equal with respect to their gene expression phenotype. Under this assumption, a mutation that transforms two such circuits into one another would be neutral with respect to this phenotype (Fig.~1b).

\subsection*{Determination of 1-mutant neighborhoods}

In several of our analyses, we explored properties of the circuits that differ from a reference circuit genotype $G$ by one single mutation. For each entry $a_{ij}$ in the matrix ${\mathbf A}$ of $G$ we considered the following 
cases: i) if $a_{ij}=0$ we check the phenotype of two mutants, one in which $a_{ij} <0$, and one in which $a_{ij} > 0$; ii) if $a_{ij}\neq 0$ we also check the phenotype of two mutants, one in which an interaction is lost ($a_{ij}=0$), and one in which we change 
the value of $a_{ij}$ while keeping its sign unchanged. Among all the variants in the one-mutation neighborhood, we allowed exclusively those that maintained the number of interactions within an interval [$m_{-}, m_{+}$], thus keeping interaction density at a value close to $c$. In all cases $m_{+}-m_{-}=5$. We discarded circuits that did not attain a steady-state gene activity pattern in this approach. Whenever a new non-zero value was required for a given $a_{ij}$, we chose an N(0,1) pseudorandom number, and forced its sign if needed. We defined the robustness to mutations of a genotype $G$ as the fraction of $G$'s 1-mutant neighbors that produce the same phenotype as $G$ when their dynamics start from the initial state $s_0$. To assess the phenotypes that $G$ can access through mutations we registered and counted all the different phenotypes produced by the set of single mutant circuits that neighbor the reference circuit $G$. The approach is readily extended to entire populations, where we determined all the different phenotypes that occur in the neighborhood circuits in the population. We counted phenotypes that occurred in the neighborhood of two or more circuits only once. 

\subsection*{Evolving populations}

For the model we use, a given pair of initial and final expression states ($s_0$, $s_\infty^{opt}$) is representative of all pairs with the same fraction $d$ of individual genes' expression values that differ between $s_0$ and $s_\infty^{opt}$. For a pre-specified $d$, we thus chose an arbitrary such pair, and followed previously established procedures \citep{Cilibertietal2007} to identify a circuit genotype $G$ that is able to drive the system from $s_0$ to $s_\infty^{opt}$. The regulatory interactions in the initial genotype $G$ are real numbers sampled from a normal distribution with mean 0 and standard deviation 1, i.e., an N(0,1) distribution. After having identified one such genotype $G$, we created a population of 200 copies of it, and subjected this population to repeated cycles (``generations'') of mutations (with a probability of mutation of $\mu=0.5$ per circuit), and strong stabilizing selection on $s_\infty^{opt}$. Whenever a circuit underwent mutation, we picked one of the circuit's 1-mutation neighbors at random (see above). We allowed exclusively those mutations that maintained the number of interactions within a predetermined interval [$m_{-}, m_{+}$], thus keeping interaction density around a predetermined fraction $c$ of non-zero interactions among all possible ($N^2$) interactions. In this study, $m_{+}-m_{-}=5$ in all cases.

Throughout, we interpret a circuit's ``fitness'' as a survival probability. We followed the regulatory dynamics of each gene circuit with $s_0$ as initial condition. As in previous work \citep{Cilibertietal2007pnas,Cilibertietal2007}, we disregarded genotypes that did not produce fixed-point equilibrium states, or that produced phenotypes in which the activity of a gene was equal to zero (neither active nor inactive). We assigned genotypes that attained the reference pattern $s_\infty^{opt}$ a maximally possible fitness of 1. Thus, $s_\infty^{opt}$ represents a pre-determined optimal gene expression state. We assigned genotypes that attained an equilibrium state $s_\infty$ that differed from $s_\infty^{opt}$ in the activity state of $k$ genes a fitness equal to $(1 - k/N)^5$, which ensures a steep decrease in survival probability even for small deviations from $s_\infty^{opt}$. Each generation, we constructed a new population by sampling individuals with replacement from the previous generation, and subjecting copies of them to mutation with a probability $\mu$. We kept each of these new individuals with a probability equal to its fitness, and continued sampling until the newly generated population had 200 members. For all the populations we study, we let the initial population of identical genotypes evolve for $10^4$ generations under selection for $s_\infty^{opt}$, before collecting any simulation data. This allows the population to erase any traces of the initial genotype and to reach a plateau where phenotypic variability in response to either mutations or non-genetic perturbations varies little across generations.

We define the genotypic distance between two circuits as the minimum number of mutations needed to transform one circuit into the other, normalized by the maximally possible number of such mutations for circuits with the same number of interactions. 

\subsection*{Noisy dynamics}
For each circuit in a population, we generated $5N$ dynamic trajectories, each of which started from $s_0$. For each of these trajectories, and for each step of the regulatory dynamics, we perturbed the activity of a randomly picked gene with a probability of 0.5. We then followed each trajectory until an activity pattern $s$ had consecutively repeated itself, and labeled this pattern as $s_{\infty}$. We then counted the number of different fixed-point equilibrium states that each circuit could attain in these $5N$ trajectories. 

\begin{figurehere}
\centering
\resizebox{\columnwidth}{!}{\includegraphics{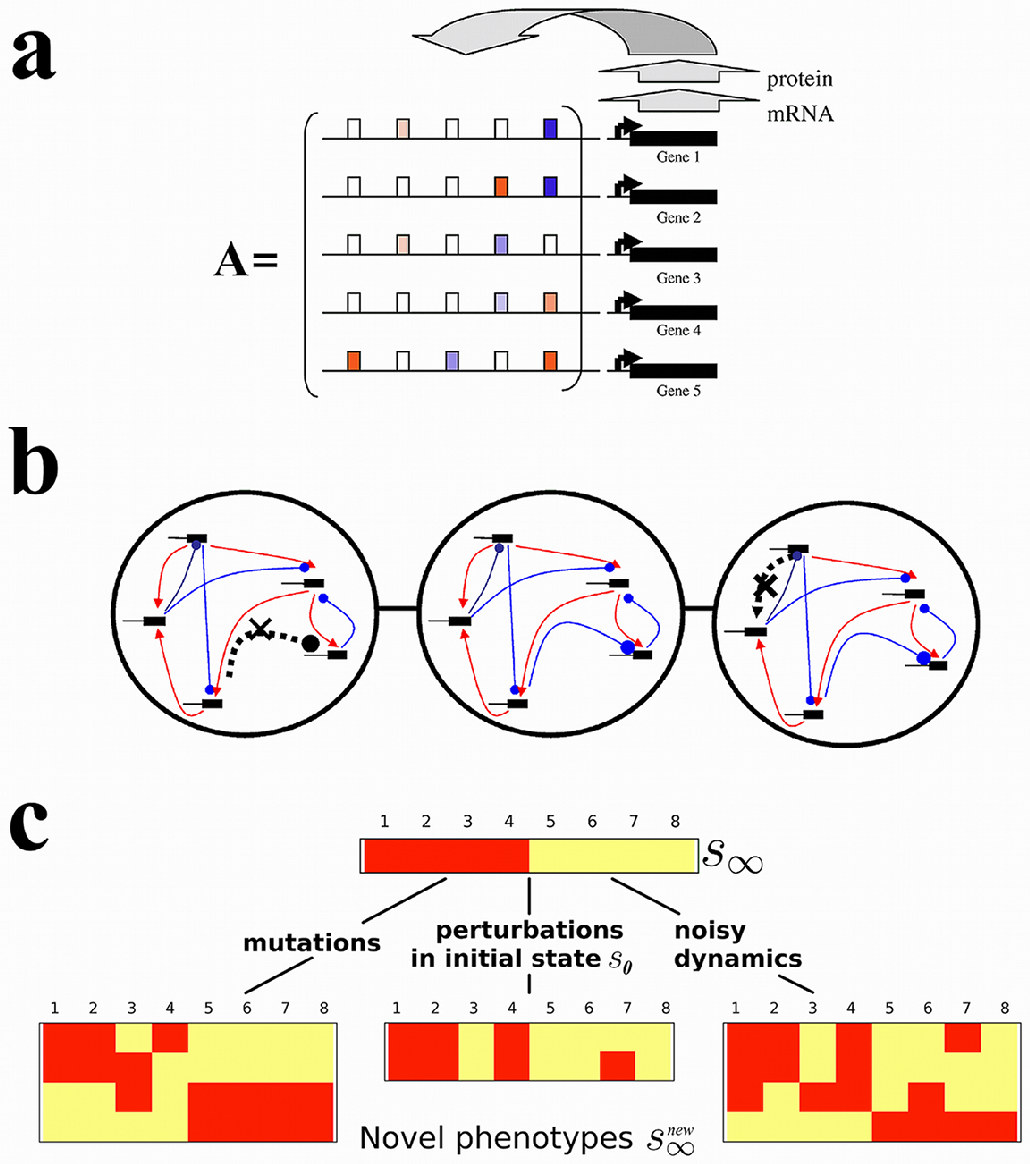}}
\caption{Gene regulatory circuit model. (a) A gene regulatory circuit. Black bars indicate genes that encode proteins which regulate the activity of other genes in a hypothetical circuit. The regulatory interactions are described by a matrix ${\mathbf A} = (a_{ij})$. An interaction means that the activity of gene $j$ can either have a positive ($a_{ij} > 0$, red rectangles) or a negative ($a_{ij} < 0$, blue rectangles) effect on the activity of gene $i$. (b) Gene circuits that differ in a single interaction are neighbors in genotype space. . Each large circle surrounds a distinct gene regulatory circuit. Red arrows represent activating interactions, and blue lines represent repressing interactions between different genes (black rectangles). Dashed lines represent the interactions that are necessary to convert the indicated circuits into the middle circuit. a) and b) are modified with permission from \citep{Cilibertietal2007}. (c) Example of novel phenotypes caused by three kinds of perturbations. A reference gene circuit produces phenotype $s_\infty$, in which four genes are active (genes 5-8; yellow) and four genes are inactive (genes 1-4; red). In a one-mutation neighborhood (all circuits that differ from the reference one by a single interaction) we find four phenotypes $s_\infty^{new}$ different from the original $s_\infty$ (left). If we perturb the system state of the reference circuit without altering its genotype, other novel phenotypes are encountered (center and right panels). The perturbations we used are either all single-gene perturbations in the initial condition $s_0$ (center), or perturbations of the dynamical trajectory of the circuit (`noisy dynamics'; right). }
\end{figurehere}

\subsection*{Random sampling of genotypes in genotype networks}

In order to sample properties of a given genotype network uniformly, we performed a random mutational walk restricted to this genotype network, that is, to circuits that attain a given $s_\infty^{opt}$ from the initial state $s_0$. We then examined properties of genotypes every $n$ steps of this random walk, where $n$ equaled 5 times the upper limit $m_{+}$ of the number of interactions in the circuit. This sporadic sampling serves to erase correlations in genotypes along this random walk. 

\section*{Results}

\subsection*{Genotype networks of gene expression phenotypes have different sizes}

For our model, most or all genotypes that produce the same phenotype form large connected genotype networks \citep{Cilibertietal2007pnas,Cilibertietal2007}. The size of any one phenotype's genotype network depends only on the fraction $d$ of genes whose expression state differs between the initial state $s_0$, and the steady state activity phenotype $s_\infty^{opt}$ \citep{Cilibertietal2007}. Specifically, phenotypes where these two states (regardless of their actual expression values) are more similar have larger genotype networks (Fig. S1 in Supplementary Material). One can view regulatory circuits as devices that compute an expression state $s_\infty^{opt}$ from the initial state $s_0$. From this perspective, a larger number of gene expression differences between these states means that the computation becomes increasingly difficult, in the sense that fewer genotypes can perform it. 

We first examined, in our model, the relationship between the size of a phenotype $P$'s genotype network and the robustness of circuits with this phenotype $P$ to mutations. To this end, we uniformly sampled $10^6$ genotypes from genotype networks of different sizes (different $d$), and determined their mean robustness to mutations, that is, the mean fraction of their neighbors with the same phenotype. For all examined cases, the average mutational robustness (i.e. phenotypic robustness) is higher for genotypes on larger genotype networks (Fig. S2). Thus, phenotypic robustness to mutations increases with genotype network size, just as for RNA \citep{Wagner2008prsb}. Therefore, we can simply use $1-d$ as a proxy for genotype network size and phenotypic robustness to mutations. 

\subsection*{Phenotypic robustness to mutations facilitates phenotypic variability in response to noise} 

In this paper, we are concerned with the production of new steady-state gene expression patterns $s_\infty^{new}$ that are different from $s_\infty^{opt}$. We refer to such activity patterns as new phenotypes. They could result from mutations that change regulatory interactions in a circuit. They could also result from {\it non-genetic} perturbations (Fig.~1c). We here consider two kinds of non-genetic perturbations, noise in a cell's internal milieu, and change in the organism's (external) environment. Both kinds can induce dramatic gene expression changes in organisms ranging from bacteria to metazoans \citep{Rajetal2010,Snell-Roodetal2010,Elowitzetal2002}. We first focus on noise, which includes stochastic changes in protein or mRNA copy numbers in a cell, and can cause phenotypic heterogeneity in clonal populations \citep{McAdamsyArkin1997,Elowitzetal2002,Rajetal2010}. Such noise may affect the activity or expression of circuit genes at a given time, which may alter a circuit's gene expression dynamics, and lead to a new steady-state activity pattern $s_{\infty}^{new}$. 

We emulated the perturbations produced by noise in two complementary ways. First, we changed the activity state of single genes in the initial state $s_0$, for each gene in a circuit, and determined the different new phenotypes $s_\infty^{new}$ that resulted from such change. Secondly, we randomly perturbed the dynamic trajectory from $s_0$ to $s_\infty$ (`noisy dynamics'), as described in Methods. 

We asked how the mutational robustness of a gene expression phenotype affects the number of new phenotypes that these two kinds of noise can produce in populations of evolving circuits. We evolved populations of 200 circuits under stabilizing selection on a given gene expression state $s_\infty^{opt}$, as described in Methods. We then counted the number of unique new phenotypes that the two different kinds of noise produced among the individuals in a population. Specifically, we counted phenotypes that appeared multiple times only once, thus focusing on unique phenotypes.

Noise can produce a greater number of new phenotypes in populations evolving on large genotype networks. Fig. 2 demonstrates these observations for circuits with $N=20$ genes and an interaction density $c \approx 0.2$. These observations also hold if we vary the numbers of genes and regulatory interactions in a circuit (Figs. S3,4), with a single exception for perturbations in $s_0$ when the number of regulatory interactions is very low (Fig. S3d). 

\subsection*{Populations with more robust phenotypes harbor more diverse genotypes}
Increased genotypic diversity in populations evolving in large genotype networks might aid in producing increased phenotypic variability, as discussed in the Introduction. We next asked whether this mechanism may apply to our system. To this end, we quantified the genotypic distance among two circuits (see Methods). 

As a measure of a population's genotypic diversity, we estimated the mean pairwise circuit distance, as well as its maximum, in each of 500 populations evolved under stabilizing selection on a phenotype $s_\infty^{opt}$. We did so for two classes of populations that differ in the robustness of their phenotypes, and found that the mean genotypic distance is significantly higher for populations with a robust phenotype. The same holds also for the maximum genotypic distance. These observations are not sensitive to the number of genes and interactions in a circuit (Table S1). Thus, populations with a robust phenotype are genetically more diverse than populations with a less robust phenotype. These observations hint that the higher genetic diversity of populations with robust phenotypes may be exposed as phenotypic variability in response to noise. 

\begin{figurehere}
\centering
\resizebox{0.9\columnwidth}{!}{\includegraphics{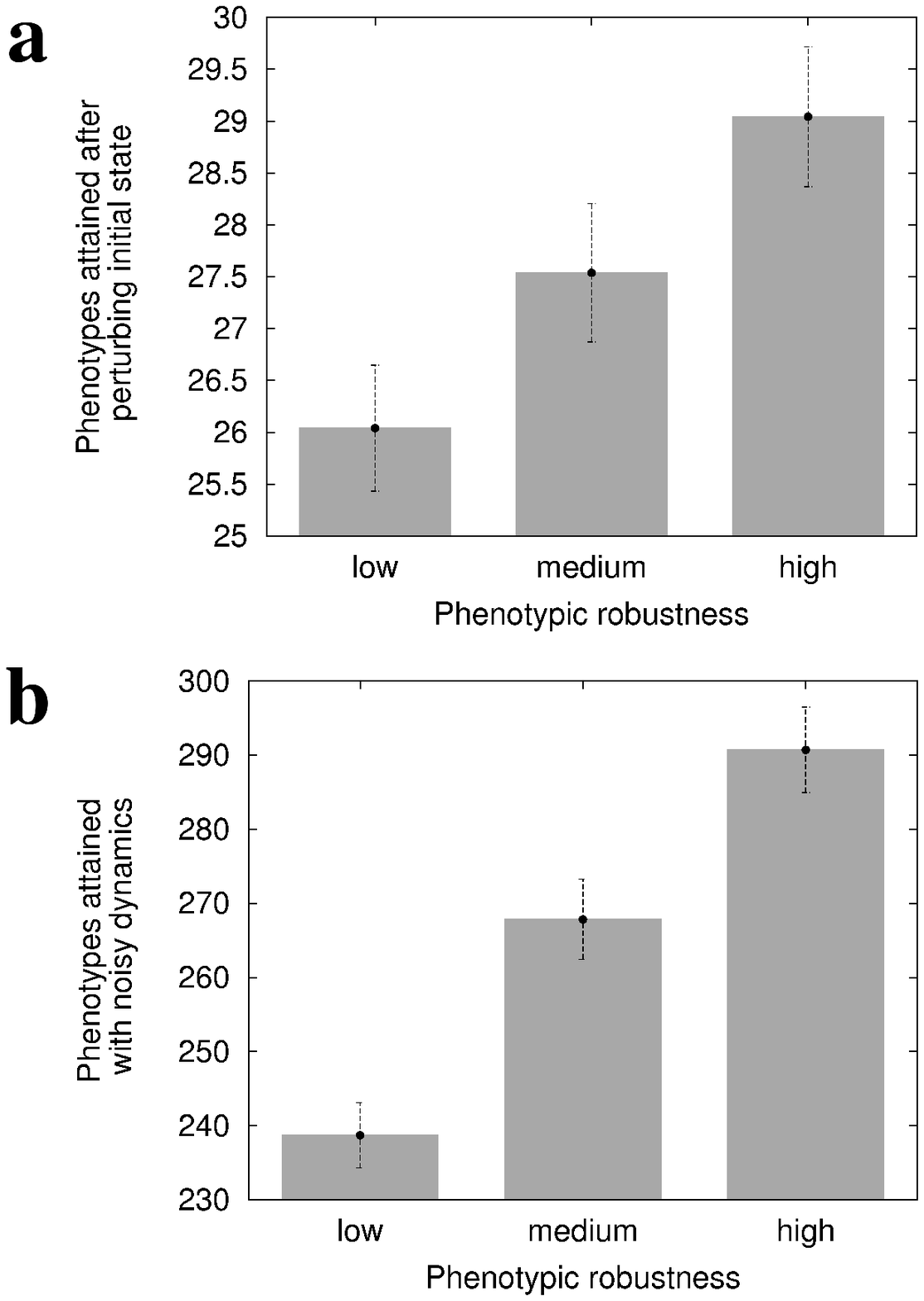}}
\caption{High phenotypic robustness facilitates phenotypic variability in response to noise in gene expression. The distance $d$ between the initial state $s_0$ and the optimal phenotype $s_\infty^{opt}$ is strongly associated with genotype network size and with phenotypic mutational robustness (``phenotypic robustness'') hereafter. `High', `medium' and `low' correspond to expression phenotypes with high ($d=0.1$), intermediate ($d=0.25$), and low ($d=0.5$) robustness. The figure shows results for $N=20$ genes and a fraction $c\approx0.2$ of non-zero regulatory interactions. Both panels show mean numbers of novel phenotypes averaged over 500 independent populations, for each level of robustness. The length of bars denotes one standard error. The number of different new phenotypes that a population can access after perturbations of a) single genes in the initial state $s_0$, or b) a circuit's gene expression trajectory, increases with phenotypic robustness.}
\end{figurehere}

\subsection*{Phenotypic robustness does not facilitate phenotypic variability caused by mutations}

 We next asked whether phenotypic robustness also facilitates phenotypic variability in response to mutations for the regulatory circuits we study. We again studied populations of circuits evolved under stabilizing selection on a phenotype $s_\infty^{opt}$. In such populations, we determined the 1-mutation neighborhood of each circuit, that is, all circuits that differ from it in a single regulatory interaction. We then determined the number of unique new gene expression phenotypes in the population's neighborhood. That is, we counted only once a phenotype if it occurred in the neighborhood of two different circuits. This number of unique phenotypes is a measure of the population's phenotypic variability in response to mutations.

\begin{figurehere}
\centering
\resizebox{\columnwidth}{!}{\includegraphics{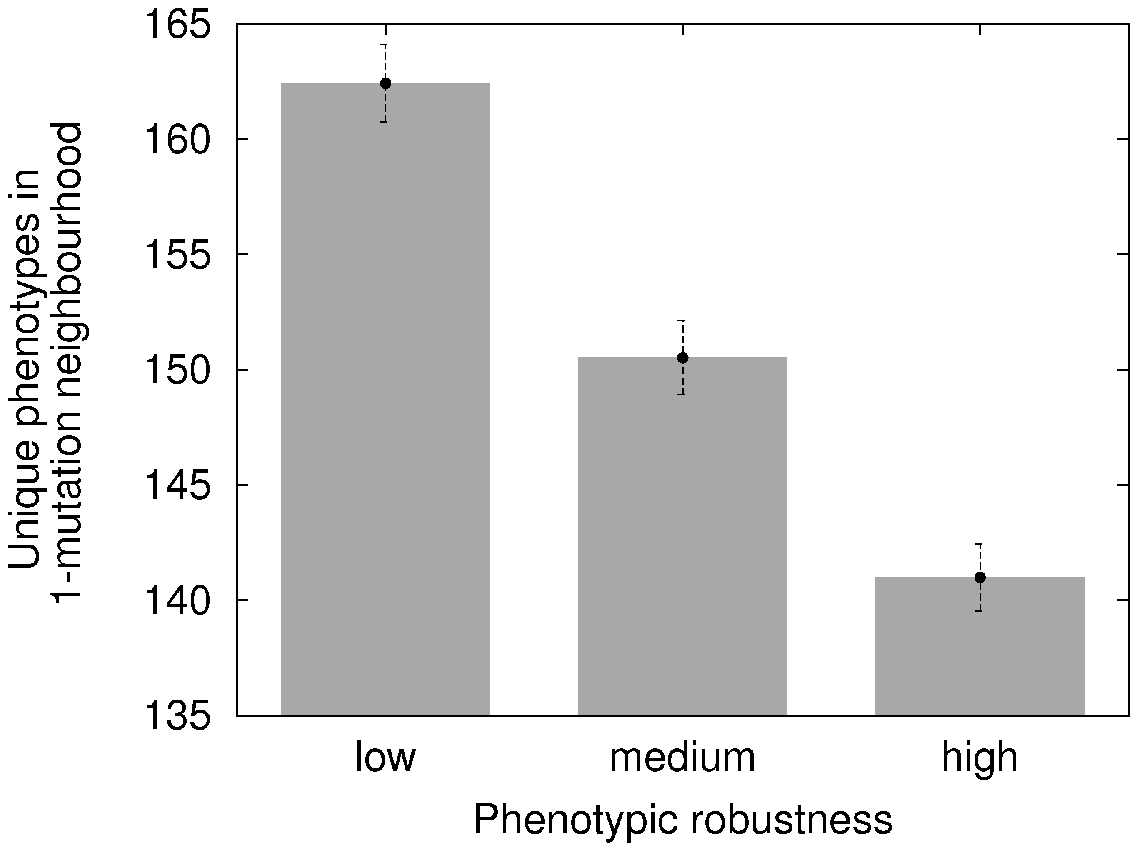}}
\caption{High phenotypic robustness does not facilitate phenotypic variability in response to mutations without preceding environmental change. `High', `medium' and `low' correspond to expression phenotypes with high ($d=0.1$), intermediate ($d=0.25$), and low ($d=0.5$) robustness. The figure shows results for $N=20$ genes and a fraction $c\approx0.2$ of non-zero regulatory interactions. The panel shows the mean number of novel phenotypes averaged over 500 independent populations, at each level of robustness. The length of bars denotes one standard error.}
\end{figurehere}

We found that populations with a highly robust phenotype show lower phenotypic variability in response to mutations. This holds despite their somewhat higher genotypic diversity (Table S1, discussed above). Fig. 3 shows pertinent data for circuits with 20 genes and interaction density $c\approx0.2$. The same behavior holds for populations of circuits with different number of genes and different interaction densities (Fig. S5). In sum, robustness of a phenotype to mutations impairs phenotypic variability to mutation, as opposed to what we saw for variability in response to noise. 

Our results suggest that a phenotype's mutational robustness promotes phenotypic variability in response to noise, but hinders such variability in response to mutations. This may seem surprising, because robustness to mutations increases with robustness to noise for individual circuits \citep{Cilibertietal2007}. One might thus think that phenotypic variability also behaves similarly in response to these perturbations. However, robustness to mutations explains less than 25 percent of the variance in robustness to noise, as shown by a new statistical analyses of our previously published data \citep{Cilibertietal2007}. Thus phenotypic variability in response to noise and to mutation are only weakly coupled.

With these observations in mind, we analyzed the phenotypic variability in response to noise and mutations of individual circuits in populations evolving on different genotype networks (Table S2). After obtaining this data, we compared the mean number of new phenotypes that mutations or noise could produce from circuits in populations with different levels of phenotypic robustness (Table S3). We found that phenotypic variability in response to gene expression noise decreases less with phenotypic robustness to mutations than phenotypic variability to mutations (Table S3). It may even increase with phenotypic robustness. These observations suggest that the increased genotypic diversity attained on larger genotype networks is insufficient to compensate for the small (or null) reduction in variability in response to mutations. It is, however, sufficient to compensate for the smaller reduction in phenotypic variability in response to noise in gene expression.

\subsection*{Phenotypic robustness increases phenotypic variability after environmental change}

Thus far, we focused mostly on phenotypic variability in response to small, random non-genetic perturbations, such as single gene expression perturbations along a gene expression trajectory. 
We now turn to the question of what happens when a whole population is subject to the same non-genetic perturbation. In nature, this may occur because of environmental change outside the organism, or colonization of a new habitat.

The environment can have two different roles in this context. The first is an inducing role, where the environment acts as an ``agent of development'' \citep{West-Eberhard1989}. In this role, it affects the phenotype produced from a genotype. In many cases, environmentally induced phenotypic change is linked to major changes in gene expression \citep{Snell-Roodetal2010}. The second role is an evaluating role, where the environment acts as an ``agent of selection'' \citep{West-Eberhard1989}. In anthropomorphic terms, the environment in this role distinguishes well-adapted from poorly adapted phenotypes. 

Conveniently, our model allows us to study these roles independently. We model a change in the environment's evaluation role as a change in the identity of the optimal phenotype $s_{\infty}^{opt}$, for all circuits in the population. We model a change in the environment's inducing role as a change in the initial state $s_{0}$ in the whole population. Such a change could occur, for example, through a signaling pathway that detects an environmental change, and that affects genes upstream of the circuit. Put differently, changes in $s_0$ reflect the environment's effect on phenotype production, while changes in $s_\infty^{opt}$ affect the survival probability of individuals, without inducing novel phenotypes. We note that other factors, such as mutations in upstream genes, might also lead to changes in $s_0$. Any one such change, however, would initially affect only one individual in a population, and not the whole population at the same time.

We first asked how an environmentally induced change in the initial gene activity pattern $s_0$ affects the number of different actual phenotypes that a population displays. We note that our populations may contain a few individuals with phenotypes different from the optimal phenotype $s_{\infty}^{opt}$. The reason is that, in contrast to previous formulations \citep{Cilibertietal2007}, we here represent fitness as a continuous variable that depends on the similarity of a circuit's phenotype $s_{\infty}$ to $s_{\infty}^{opt}$ (see Methods). We started out with a population evolved under stabilizing selection on an optimal expression phenotype $s_\infty^{opt}$ and a given gene activity pattern $s_0^a$ as initial condition. We counted the number of phenotypes in the population, and compared it with the number of different phenotypes that the same population displays when $s_0^a$ is replaced by a random gene activity pattern $s_0^b$ as an initial condition. We found that phenotypic variability increases after substitution of $s_{0}^{a}$ with $s_{0}^{b}$ (Figs.~4 and S6). In addition, the magnitude of this increment increases with phenotypic robustness (Fig.~4). 
This last observation is generally not sensitive to the number of genes and regulatory interactions in a circuit (Fig. S6). The increase in phenotypic variability is significantly higher for populations with a robust phenotype (Mann-Whitney U-test; Table S4). The single exception to these observations were circuits of very low interaction density ($N=20$; $c\approx0.1$ Fig. S6d), that also show other non-typical behaviors \citep{Cilibertietal2007}. Our results suggest that, after environmental change, observable phenotypic diversity increases to a larger extent in populations with a robust phenotype. We note that because the identity of $s_{\infty}$ does not affect the production of phenotypes, but only their viability, it is not appropriate to carry out an analogous analysis for changes in $s_\infty^{opt}$. 

In earlier sections, we have shown that phenotypic robustness impedes phenotypic variability after mutations in populations evolving in a constant environment (Fig.~3). We next asked whether this also holds after a change in the inducing role of the environment. We started out, as in our last analysis, with populations of circuits evolved under stabilizing selection on an optimal expression phenotype $s_\infty^{opt}$ and with a given gene activity pattern $s_0^a$ as initial condition. Then, we changed the initial condition $s_{0}^{a}$ for all the circuits to a new random initial condition $s_{0}^{b}$, and allowed evolution to proceed. Before and after this change, we recorded the number of different phenotypes accessible from the population through mutations. Under the new condition the population effectively searches genotype space for optimal phenotypes. During this search, many variant circuits may not pass to subsequent generations. Our primary focus, however, is not this search, but the immediate increase in a population's phenotypic variability in response to environmental change. 

Before environmental change, populations with a robust phenotype have access to fewer phenotypic variants, just as in our previous observations (Fig.~3). Immediately after environmental change (at $t=1$), the number of new phenotypes accessible through mutations increases, in a burst, in all populations. Importantly, this increase is higher in populations with a robust phenotype (Figs.~5a and S7). This means that phenotypic robustness facilitates the phenotypic variability caused by mutations, but only after environmental change. As in our analysis above, the only exception occurs when interaction density is very low (Fig. S7d). 

\begin{figurehere}
\centering
\resizebox{\columnwidth}{!}{\includegraphics{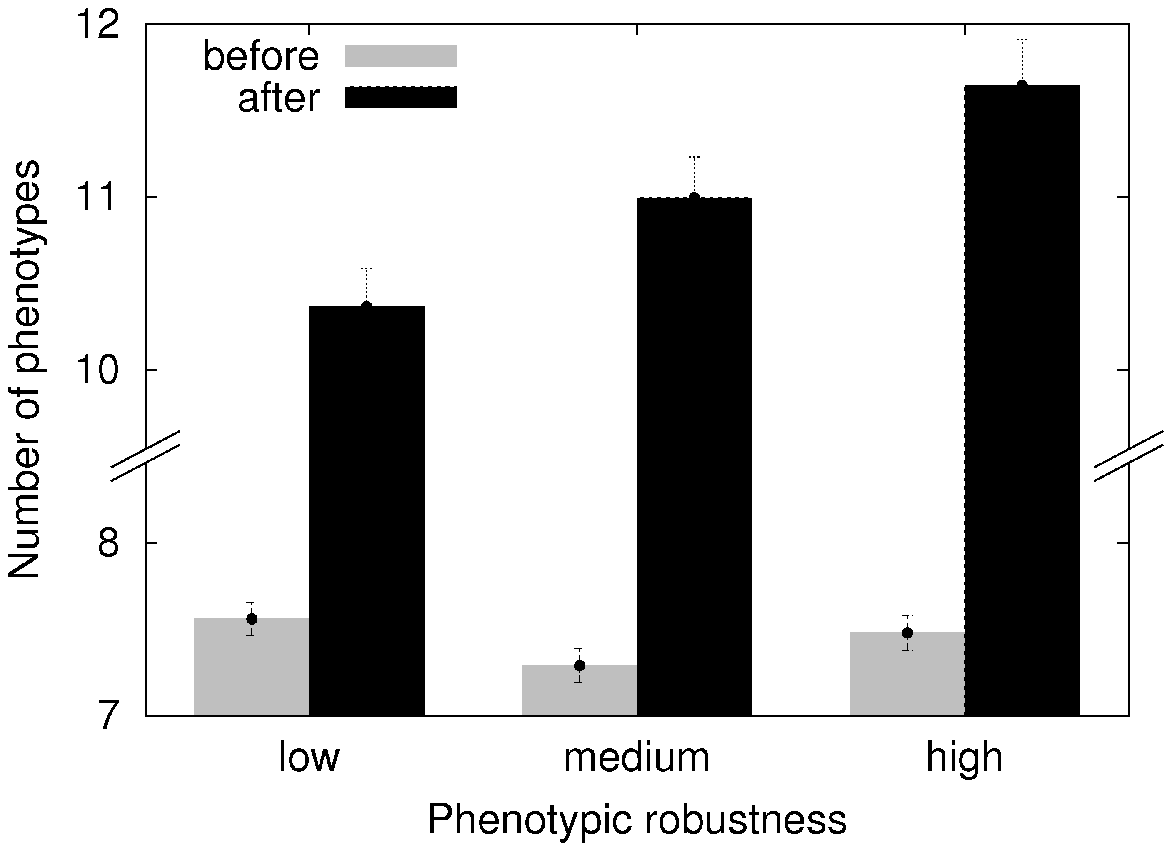}}
\caption{High phenotypic robustness increases phenotypic diversity in populations of gene circuits after environmental change. The number of different phenotypes that populations display increases after changing the initial expression state $s_0$. Such an increase is greater for populations with mutationally more robust phenotypes. The figure shows results for $N=20$ genes and a fraction $c\approx0.2$ of non-zero regulatory interactions. The panel shows the mean number of observed phenotypes averaged over 500 independent populations, for each level of robustness. The length of bars denotes one standard error.}
\end{figurehere}

We next asked whether the evaluation role of the environment has similar effects on the mutational access to new phenotypes. To this end, we repeated the above analysis, but replaced, at $t=1$, the optimal phenotype $s_{\infty}^{opt,a}$ by a randomly chosen optimal $s_{\infty}^{opt,b}$ (without changing $s_{0}$). We also observed a transient, but more gradual, increase in the number of phenotypes that are mutationally accessible. However, in this case, phenotypic variability is lower for populations with a robust phenotype after environmental change (Figs.~5b and S8). Thus, the inductive role of environment, but not its evaluative role, causes higher phenotypic variability in populations with robust phenotypes. 

An open question is how mutation-accessible phenotypic variability changes when both the inductive and the evaluation roles of the environment change. This question is important because a change in the inductive role favors phenotypic variability to a larger extent in populations with a robust phenotype, whereas a change in the evaluative role particularly favors variability in populations with less robust phenotypes. Thus, a
combination of both effects could result in a negligible effect of phenotypic robustness on variability after environmental change.
To answer this question, we repeated our analysis from the previous paragraph, but replaced the original {\it pair} of states ($s_0^a$, $s_\infty^{opt,a}$) with a new pair ($s_0^b$, $s_\infty^{opt,b}$), such that the distance $d$ between $s_0$ and $s_\infty^{opt}$ was the same for both pairs. 

In this new analysis, populations evolving on a large genotype network show greater phenotypic variability in response to mutations immediately after this change (Fig. 5c). These differences are statistically highly significant (Table S5). The same observations hold for circuits of different sizes and different connectivities (Fig. S9 and Table S5). As in our analysis above, the only exception occurs when interaction density is very low (Fig. S9d). These observations imply that the inductive role dominates in its immediate effect on phenotypic variability when both roles of the environment change. In sum, populations with a robust phenotype have mutational access to more phenotypic variants after environmental change. This increased access is caused by the inductive role of the environment, that is by the new phenotypes that a new environment can bring forth.

\section*{Discussion} 
 
If non-genetic change is to be causally involved in evolutionary innovation, it needs to generate novel phenotypes. Genetic assimilation can then stabilize these phenotypes if the non-genetic perturbations that induced them appear recurrently. Here, we focused on whether the robustness of an existing phenotype and non-genetic change can facilitate the origin of new phenotypes. To address this question, we examined a generic model of transcriptional regulation circuitry, in which the relationship between genotypes (patterns of regulatory interactions) and phenotypes (gene activity or expression patterns) is well-studied \citep{Wagner1996,Cilibertietal2007pnas,Cilibertietal2007,MartinyWagner2008,MartinyWagner2009}. In this model, we can use the size of a phenotype's genotype network as a proxy for a phenotype's robustness to mutations. 

We broadly distinguished two kinds of non-genetic perturbations. The first corresponds to fluctuations in a gene circuit's microenvironment that produce random changes in gene expression. Such changes, are ubiquitous and have important effects on cell biological processes \citep{McAdamsyArkin1997,Elowitzetal2002,Rajetal2010}. 
The second kind comprises changes in the (macro)environment external to an organism. For brevity, we refer to these kinds of change as noise and environmental change. 

We first explored how noise and mutations affect phenotypic variability in populations that differ in the robustness of their gene activity phenotypes. We found that phenotypic mutational robustness increases phenotypic variability of populations in response to noise but not in response to mutations. This last finding differs from observations for RNA secondary structure, where phenotypic robustness facilitates the mutational access to phenotypic variants \citep{Wagner2008prsb}. The reason stems from differences in the organization of genotype space for these two system classes, i.e. in the distribution of different genotype networks in genotype space \citep{Wagner2008prsb,Cilibertietal2007pnas,Espinosa-Sotoetal2010,AncelyFontana2000}. A recent mathematical model \citep{Draghietal2010} shows that mutational access to new phenotypes can depend on this organization, and on details of a population's evolutionary dynamics.

\begin{figure*}
\centering
\includegraphics[width=\textwidth]{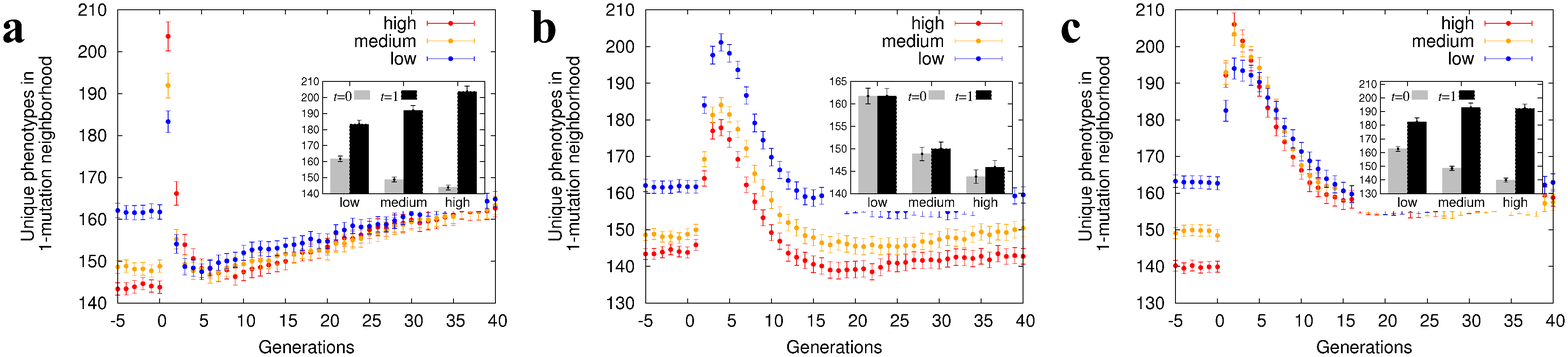}
\caption{High phenotypic robustness allows mutational access to more phenotypes after an environmental change produces novel phenotypes.The figure shows the number of different phenotypes in the 1-mutant neighborhood of a population, for three different scenarios of environmental change at generation $t=1$. The insets show the number of phenotypes accessible through mutation immediately before ($t=0$) and immediately after ($t=1$) environmental change. The figure shows results for $N=20$ genes and a fraction $c\approx0.2$ of non-zero regulatory interactions. `High', `medium' and `low' correspond to expression phenotypes with high ($d=0.1$), intermediate ($d=0.25$), and low ($d=0.5$) phenotypic robustness. Data are mean values averaged across 500 independent simulations for each level of robustness. The length of bars denotes one standard error. (a) The initial state $s_0$ changes. (b) The identity of the optimal phenotype $s_\infty^{opt}$ changes. (c) The original pair ($s_0^a$, $s_\infty^{opt,a}$) changes to a new pair ($s_0^b$, $s_\infty^{opt,b}$).}
\end{figure*}

Next, we asked how phenotypic robustness affects phenotypic variability in response to environmental change, which we modeled to affect all individuals in a population. We showed that environmental change, besides increasing observable phenotypic variation, transiently increases phenotypic variability caused by mutations. Because mutational access to most novel variants is only possible in the new environment, these variants can be considered environmentally-induced phenotypes, supporting the phenotype-first scenario. From this perspective, environmental change increases phenotypic variability and the chances to refine, exaggerate and stabilize phenotypic variation via genetic change. Importantly, this increase in phenotypic variability is higher in populations that had a more robust phenotype before environmental change. 

In sum, we found a positive effect of phenotypic robustness on phenotypic variability after non-genetic perturbations. This positive effect is not sensitive to the magnitude of non-genetic perturbations. Phenotypic robustness favors phenotypic variability after single-gene perturbations in the initial condition, which is the smallest possible non-genetic perturbation in our model (Fig.~2a). It also favors phenotypic variability after replacing the initial condition by a completely new random activity pattern (Figs.~4 and 5a).
In contrast, populations with robust gene expression phenotypes produce less phenotypic variation when only mutations are used to explore the phenotypic possibilities. Thus, a mechanism that relies exclusively on mutation to produce novel phenotypes becomes less important for innovation as a phenotype's robustness increases. Our results suggest that plasticity-mediated innovation may be especially important for gene expression traits with high mutational robustness. Our work is an initial step towards the definition of two different domains where either the genotype-first or phenotype-first scenarios prevail over the other. In this regard, we note that environmental induction of novel traits is not only expected for gene circuits with robust phenotypes. In fact, we observe a general increase in phenotypic variability after environmental change (Figs.~4 and~5). It is just that this increase is especially strong for robust phenotypes. 

The general increase in phenotypic variability we observe is consistent with many empirical observations on phenotypic variation that is conditional on the environment. For example, severe environments enhance phenotypic differences among fruit fly strains \citep{KondrashovyHoule1994}, and a temperature rise due to an unshaded milieu increases the frequency of abnormal morphologies in fruit flies \citep{RobertsyFeder1999}. Moreover, population genetics studies show that the release of hidden genetic variation after environmental change should be very common 
\citep{HermissonyWagner2004}. 

Unfortunately, because the mutational robustness of most traits is unknown and difficult to measure \citep{deVisseretal2003}, these observations do not speak directly to the question whether phenotypic robustness promotes phenotypic variability after environmental change. However, there is an intriguing phenomenon with potential parallels to our observations. In the fruit fly {\it D. melanogaster} and in the plant {\it Arabidopsis thaliana}, previously unobserved phenotypic variants appear when the activity of the chaperone protein Hsp90 is impaired (by environmental stress or by other means) \citep{RutherfordyLindquist1998,Queitschetal2002}. This observation suggests that an active Hsp90 protein usually conceals genetic variation that would otherwise be visible phenotypically, although the appearance of new genetic variants may underlie some of the new phenotypic variation \citep{Specchiaetal2010}. Hence, Hsp90 might increase phenotypic variability by allowing the accumulation of cryptic genetic variation that can become phenotypically visible in the right environment. Our analysis revealed an analogous phenomenon for robust gene expression phenotypes. Robustness of phenotypic traits can facilitate the accumulation of cryptic genetic variation in a population. Non-genetic perturbations can expose this variation as new phenotypes, some of which may be adaptive. The parallels between the action of Hsp90 and the phenomenon we study may be more than superficial. The reason is that Hsp90 increases the number of genotypes with a given phenotype for several traits \citep{RutherfordyLindquist1998,Rutherford2003}. It thus increases phenotypic robustness. Hsp90 is a specific example -- with a peculiar mechanism -- of how phenotypic robustness can increase phenotypic variability after non-genetic perturbations. 

In conclusion, our observations suggest that phenotypic robustness to mutations can play a positive role in the phenotypic variability after non-genetic perturbations. To see this, one needs to study the role of population level processes, as we did. We caution that we made our observation in the context of a specific model of transcriptional regulation circuits. The gene expression phenotypes of such circuits play central roles in many evolutionary innovations \citep{Shubinetal2009,DavidsonyErwin2006}. 
However, phenotypes may be distributed differently in genotype space in other classes of biological systems. Thus, whether our observations hold in other systems remains to be seen.

\section*{Acknowledgements}
 {\small
The research leading to these results has received funding from the European Community's Seventh Framework Programme FP7/2007-2013 under grant agreement n$^o$ PIIF-GA-2008-220274. AW acknowledges support through SNF grants 315200-116814, 315200-119697, and 315230-129708, as well as through the YeastX project of SystemsX.ch , and the University Priority Research Program in systems biology at the University of Zurich.
}

\bibliographystyle{apalike} 
{\footnotesize

}
\end{multicols}

\end{document}